\documentclass[onecolumn,resetfootnote]{aastex7}
\usepackage{xcolor}

\usepackage{natbib}
\usepackage{amsmath}
\usepackage{enumitem}

\shortauthors{Bernstein et al.}
\shorttitle{Dimensional reduction for sampled priors}
\reportnum{DES-2025-0886}
\reportnum{FERMILAB-PUB-25-0362-PPD}
\published{1 May 2026, \apj}

\newcommand{\ie}{\textit{i.e.}}
\newcommand{\eg}{\textit{e.g.}}
\newcommand{\E}{\mathrm{E}}
\newcommand{\eqq}[1]{Equation~(\ref{#1})}
\newcommand{\vecc}{\ensuremath{\mathbf{c}}}
\newcommand{\vecq}{\ensuremath{\mathbf{q}}}

\newcommand{\vecn}{\ensuremath{\mathbf{n}}}
\newcommand{\vecu}{\ensuremath{\mathbf{u}}}
\newcommand{\vecU}{\ensuremath{\mathbf{U}}}
\newcommand{\vecv}{\ensuremath{\mathbf{v}}}
\newcommand{\hatc}{\ensuremath{\bar{\mathbf{c}}}}

\newcommand{\covm}{C}

\newcommand{\matD}{D}
\newcommand{\matE}{E}
\newcommand{\matF}{F}
\newcommand{\matG}{G}
\newcommand{\matI}{I}
\newcommand{\matT}{T}
\newcommand{\matX}{X}
\newcommand{\matY}{Y}
\newcommand{\matV}{V}
\newcommand{\matLam}{\Lambda}
\newcommand{\proj}{P}  

\newcommand{\likeli}{\mathcal{L}}
\newcommand{\trace}{\text{Tr}}

\begin{document}

\title{Dimensional reduction for sampled priors and application to
photometric redshift distributions}


\suppressAffiliations

\author[0000-0002-8613-8259, gname='Gary', sname='Bernstein']{G.~M.~Bernstein}
\affiliation{Department of Physics and Astronomy, University of Pennsylvania, Philadelphia, PA 19104, USA}
\email{garyb@physics.upenn.edu}

\author[0000-0002-9719-1717, gname='William', sname='d'Assignies']{W.~d'Assignies}
\affiliation{Institut de F\'{\i}sica d'Altes Energies (IFAE), The Barcelona Institute of Science and Technology, Campus UAB, 08193 Bellaterra (Barcelona) Spain}
\email{wdoumerg@igfae.es}

\author[0000-0002-5622-5212, gname='Michael A.', sname='Troxel']{M.~A.~Troxel}
\affiliation{Department of Physics, Duke University Durham, NC 27708, USA}
\email{michael.a.troxel@gmail.com}

\author[0000-0001-8505-1269, gname='Alex', sname='Alarcon']{A.~Alarcon}
\affiliation{Institute of Space Sciences (ICE, CSIC),  Campus UAB, Carrer de Can Magrans, s/n,  08193 Barcelona, Spain}
\email{alexalarcongonzalez@gmail.com}

\author[0000-0002-6445-0559, gname='Alexandra', sname='Amon']{A.~Amon}
\affiliation{Department of Astrophysical Sciences, Princeton University, Peyton Hall, Princeton, NJ 08544, USA}
\email{alexandra.amon@princeton.edu}

\author[0000-0002-3730-1750, gname='Giulia', sname='Giannini']{G.~Giannini}
\affiliation{Kavli Institute for Cosmological Physics, University of Chicago, Chicago, IL 60637, USA}
\email{giulia.giannini15@gmail.com}

\author[0009-0006-5604-9980, gname='Boyan', sname='Yin']{B.~Yin}
\affiliation{Department of Physics, Duke University Durham, NC 27708, USA}
\email{by84@duke.edu}

\author[0000-0001-5679-6747, gname=Michel, sname=Aguena]{M.~Aguena}
\affiliation{Laborat\'orio Interinstitucional de e-Astronomia - LIneA, Av. Pastor Martin Luther King Jr, 126 Del Castilho, Nova Am\'erica Offices, Torre 3000/sala 817 CEP: 20765-000, Brazil}
\email{aguena@linea.org.br}

\author[0000-0002-7069-7857, gname=Sahar, sname=Allam]{S.~S.~Allam}
\affiliation{Fermi National Accelerator Laboratory, P. O. Box 500, Batavia, IL 60510, USA}
\email{sallamatfnalgov@gmail.com}

\author[0000-0003-0171-6900, gname='Felipe', sname='Andrade-Oliveira']{F.~Andrade-Oliveira}
\affiliation{Physik-Institut, University of Zürich, Winterthurerstrasse 190, CH-8057 Zürich, Switzerland}
\email{felipeaoli@gmail.com}

\author[0000-0002-8458-5047, gname='David', sname='Brooks']{D.~Brooks}
\affiliation{Department of Physics \& Astronomy, University College London, Gower Street, London, WC1E 6BT, UK}
\email{david.brooks@ucl.ac.uk}

\author[0000-0003-3044-5150, gname='Aurelio', sname='Carnero Rosell']{A.~Carnero~Rosell}
\affiliation{Instituto de Astrofisica de Canarias, E-38205 La Laguna, Tenerife, Spain}
\affiliation{Laborat\'orio Interinstitucional de e-Astronomia - LIneA, Av. Pastor Martin Luther King Jr, 126 Del Castilho, Nova Am\'erica Offices, Torre 3000/sala 817 CEP: 20765-000, Brazil}
\affiliation{Universidad de La Laguna, Dpto. Astrofísica, E-38206 La Laguna, Tenerife, Spain}
\email{aurelio.crosell@gmail.com}

\author[0000-0002-3130-0204, gname='Jorge', sname='Carretero']{J.~Carretero}
\affiliation{Institut de F\'{\i}sica d'Altes Energies (IFAE), The Barcelona Institute of Science and Technology, Campus UAB, 08193 Bellaterra (Barcelona) Spain}
\email{jorgecarreteropalacios@gmail.com}

\author[0000-0002-7731-277X, gname='Luiz', sname='da Costa']{L.~N.~da Costa}
\affiliation{Laborat\'orio Interinstitucional de e-Astronomia - LIneA, Av. Pastor Martin Luther King Jr, 126 Del Castilho, Nova Am\'erica Offices, Torre 3000/sala 817 CEP: 20765-000, Brazil}
\email{ldacosta@fnal.gov}

\author[0000-0000-0000-0000,gname='Maria Elidaiana', sname='da Silva Pereira']{M.~E.~S.~Pereira}
\affiliation{Hamburger Sternwarte, Universit\"{a}t Hamburg, Gojenbergsweg 112, 21029 Hamburg, Germany}
\email{elidaiana.sp@gmail.com}

\author[0000-0001-8318-6813, gname='Juan', sname='De Vicente']{J.~De~Vicente}
\affiliation{Centro de Investigaciones Energ\'eticas, Medioambientales y Tecnol\'ogicas (CIEMAT), Madrid, Spain}
\email{juan.vicente@ciemat.es}

\author[0000-0000-0000-0000,gname='Spencer', sname='Everett']{S.~Everett}
\affiliation{California Institute of Technology, 1200 East California Blvd, MC 249-17, Pasadena, CA 91125, USA}
\email{spencer.w.everett@jpl.nasa.gov}

\author[0000-0003-4079-3263, gname='Josh', sname='Frieman']{J.~Frieman}
\affiliation{Department of Astronomy and Astrophysics, University of Chicago, Chicago, IL 60637, USA}
\affiliation{Fermi National Accelerator Laboratory, P. O. Box 500, Batavia, IL 60510, USA}
\affiliation{Kavli Institute for Cosmological Physics, University of Chicago, Chicago, IL 60637, USA}
\email{frieman@fnal.gov}

\author[0000-0002-9370-8360, gname='Juan', sname='Garcia-Bellido']{J.~Garc\'ia-Bellido}
\affiliation{Instituto de Fisica Teorica UAM/CSIC, Universidad Autonoma de Madrid, 28049 Madrid, Spain}
\email{juan.garciabellido@uam.es}

\author[0000-0003-3270-7644, gname='Daniel', sname='Gruen']{D.~Gruen}
\affiliation{University Observatory, Faculty of Physics, Ludwig-Maximilians-Universit\"at, Scheinerstr. 1, 81679 Munich, Germany}
\email{dgruen@usm.uni-muenchen.de}

\author[0000-0003-2071-9349, gname='Samuel', sname='Hinton']{S.~R.~Hinton}
\affiliation{School of Mathematics and Physics, University of Queensland,  Brisbane, QLD 4072, Australia}
\email{samuelreay@gmail.com}

\author[0000-0002-9369-4157, gname='Devon L.', sname='Hollowood']{D.~L.~Hollowood}
\affiliation{Santa Cruz Institute for Particle Physics, Santa Cruz, CA 95064, USA}
\email{dhollowo@ucsc.edu}

\author[0000-0002-6550-2023, gname='Klaus', sname='Honscheid']{K.~Honscheid}
\affiliation{Center for Cosmology and Astro-Particle Physics, The Ohio State University, Columbus, OH 43210, USA}
\affiliation{Department of Physics, The Ohio State University, Columbus, OH 43210, USA}
\email{kh@fnal.gov}

\author[0000-0001-5160-4486, gname='David', sname='James']{D.~J.~James}
\affiliation{Center for Astrophysics $\vert$ Harvard \& Smithsonian, 60 Garden Street, Cambridge, MA 02138, USA}
\email{djames44@gmail.com}

\author[0000-0002-8289-740X, gname='Sujeong', sname='Lee']{S.~Lee}
\affiliation{Jet Propulsion Laboratory, California Institute of Technology, 4800 Oak Grove Dr., Pasadena, CA 91109, USA}
\email{sujeong.lee@jpl.nasa.gov}

\author[0000-0003-0710-9474, gname='Jennifer', sname='Marshall']{J.~L.~Marshall}
\affiliation{George P. and Cynthia Woods Mitchell Institute for Fundamental Physics and Astronomy, and Department of Physics and Astronomy, Texas A\&M University, College Station, TX 77843,  USA}
\email{jlm076@tamu.edu}

\author[0000-0001-9497-7266, gname='Juan', sname='Mena-Fernández']{J. Mena-Fern{\'a}ndez}
\affiliation{Universit\'e Grenoble Alpes, CNRS, LPSC-IN2P3, 38000 Grenoble, France}
\email{juan.menafernandez@lpsc.in2p3.fr}

\author[0000-0002-6610-4836, gname='Ramon', sname='Miquel']{R.~Miquel}
\affiliation{Instituci\'o Catalana de Recerca i Estudis Avan\c{c}ats, E-08010 Barcelona, Spain}
\affiliation{Institut de F\'{\i}sica d'Altes Energies (IFAE), The Barcelona Institute of Science and Technology, Campus UAB, 08193 Bellaterra (Barcelona) Spain}
\email{rmiquel@ifae.es}

\author[0000-0002-2598-0514, gname='Andrés', sname='Plazas Malagón']{A.~A.~Plazas~Malag\'on}
\affiliation{Kavli Institute for Particle Astrophysics \& Cosmology, P. O. Box 2450, Stanford University, Stanford, CA 94305, USA}
\affiliation{SLAC National Accelerator Laboratory, Menlo Park, CA 94025, USA}
\email{plazasmalagon@gmail.com}

\author[0000-0002-9646-8198, gname='Eusebio', sname='Sanchez']{E.~Sanchez}
\affiliation{Centro de Investigaciones Energ\'eticas, Medioambientales y Tecnol\'ogicas (CIEMAT), Madrid, Spain}
\email{eusebio.sanchez@ciemat.es}

\author[0000-0003-3054-7907, gname='David', sname='Sanchez Cid']{D.~Sanchez Cid}
\affiliation{Centro de Investigaciones Energ\'eticas, Medioambientales y Tecnol\'ogicas (CIEMAT), Madrid, Spain}
\affiliation{Physik-Institut, University of Zürich, Winterthurerstrasse 190, CH-8057 Zürich, Switzerland}
\email{david.sanchez@ciemat.es}

\author[0000-0002-1831-1953, gname='Ignacio', sname='Sevilla']{I.~Sevilla-Noarbe}
\affiliation{Centro de Investigaciones Energ\'eticas, Medioambientales y Tecnol\'ogicas (CIEMAT), Madrid, Spain}
\email{nsevilla@fnal.gov}

\author[0000-0000-0000-0000,gname='Tae-hyeon', sname='Shin']{T.~Shin}
\affiliation{Department of Physics and Astronomy, Stony Brook University, Stony Brook, NY 11794, USA}
\email{babohahaman@gmail.com}

\author[0000-0002-3321-1432, gname='Mathew', sname='Smith']{M.~Smith}
\affiliation{Physics Department, Lancaster University, Lancaster, LA1 4YB, UK}
\email{mat.smith@soton.ac.uk}

\author[0000-0002-7047-9358, gname='Eric', sname='Suchyta']{E.~Suchyta}
\affiliation{Computer Science and Mathematics Division, Oak Ridge National Laboratory, Oak Ridge, TN 37831}
\email{eric.d.suchyta@gmail.com}

\author[0000-0000-0000-0000,gname='Molly', sname='Swanson']{M.~E.~C.~Swanson}
\affiliation{Center for Astrophysical Surveys, National Center for Supercomputing Applications, 1205 West Clark St., Urbana, IL 61801, USA}
\email{molly.swanson@gmail.com}

\author[0000-0001-9382-5199, gname='Noah', sname='Weaverdyck']{N.~Weaverdyck}
\affiliation{Department of Astronomy, University of California, Berkeley,  501 Campbell Hall, Berkeley, CA 94720, USA}
\affiliation{Lawrence Berkeley National Laboratory, 1 Cyclotron Road, Berkeley, CA 94720, USA}
\email{nweaverd@umich.edu}

\author[0000-0002-8282-2010, gname='Jochen', sname='Weller']{J.~Weller}
\affiliation{Max Planck Institute for Extraterrestrial Physics, Giessenbachstrasse, 85748 Garching, Germany}
\affiliation{Universit\"ats-Sternwarte, Fakult\"at f\"ur Physik, Ludwig-Maximilians Universit\"at M\"unchen, Scheinerstr. 1, 81679 M\"unchen, Germany}
\email{jochen.weller@lmu.de}

\author[0000-0002-3073-1512, gname='Philip', sname='Wiseman']{P.~Wiseman}
\affiliation{School of Physics and Astronomy, University of Southampton,  Southampton, SO17 1BJ, UK}
\email{P.S.Wiseman@soton.ac.uk}

\collaboration{7}{(DES Collaboration)}


\begin{abstract}
A typical Bayesian inference on the values of some parameters of
interest $\vecq$ from some data $D$ involves running a Markov Chain (MC) to sample
from the posterior $p(\vecq,\vecn | D) \propto \likeli(D | \vecq,\vecn)
p(\vecq) p(\vecn),$ where $\vecn$ are some nuisance parameters with
  separable prior.  In some
cases, the nuisance parameters are high-dimensional, and their
prior $p(\vecn)$ is itself defined only by a set of samples that have
been drawn from some other MC.  
The MC for the posterior will typically require evaluation of
$p(\vecn)$ at arbitrary values of $\vecn,$ \ie\ one needs to provide a
density estimator over the full $\vecn$ space from the provided
samples.  But the high dimensionality of $\vecn$ hinders both the
density estimation and the efficiency of the MC for the posterior.  We
describe a solution to this problem: a linear compression of the
$\vecn$ space into a much lower-dimensional space $\vecu$ which
projects away directions in $\vecn$ space that cannot appreciably
alter $\likeli.$ The algorithm for doing so is a slight modification
to principal components analysis, and is less restrictive on
$p(\vecn)$ than other proposed solutions to this issue.
We demonstrate this ``mode projection'' technique using the analysis
of 2-point correlation functions of weak lensing fields and galaxy
density in the \textit{Dark Energy Survey}, where $\vecn$ is a binned representation of the redshift distribution
$n(z)$ of the galaxies.
\end{abstract}

\section{Motivation} \label{sec:intro}

Consider an inference in which we have a vector of observable summary
statistics \vecc\ that we are using to constrain a set of parameters of
interest \vecq.  There is a model $\hatc(\vecq,\vecn)$ for the
expectation value of the observables which involves the parameters of interest, but also a
vector \vecn\ of nuisance parameters.  
We wish to characterize the Bayesian posterior density
\begin{equation}
  p(\vecq | \vecc) \propto \int dn\, \likeli(\vecc | \vecq, \vecn) p(\vecq) p(\vecn),
\label{eq:posterior}
\end{equation}
where $\likeli(\vecc | \vecq, \vecn)$ is a known likelihood function
of the data, and we assume that the priors on $\vecq$ and $\vecn$ are separable
to $p(\vecq)$ and $p(\vecn)$.\footnote{Following the nomenclature for the
elements of Bayes's formula in, \eg, \url{https://en.wikipedia.org/wiki/Posterior_probability}.}
This posterior is complex enough that it requires
approximation by the output of a Markov Chain (MC) wandering across the space $(\vecq,\vecn).$
Implicit in \eqq{eq:posterior} is that the prior $p(\vecn)$ is independent of
  the likelihood $\likeli$ for $\vecc,$ \eg\ the prior on $\vecn$ has been constrained using
  data that are distinct from those entering the likelihood.

The scenario we address here is when \emph{the prior $p(\vecn)$ is not
    available in evaluable form, rather we have only
  a set of samples of $\vecn$ known to be drawn from this
  distribution.} Most MC samplers require that the posterior (and
hence the prior and likelihood) be an
evaluable function of any value of the parameters. It is the general task of density estimators to convert the samples of $\vecn$ into an evaluable $p(\vecn).$  But when \vecn\ is of high dimension, two problems arise: first, there may be insufficient available samples to create a viable density estimator; second, sampling of the posterior in (\ref{eq:posterior}) becomes infeasible if the MC must traverse a high-dimensional space.

A concrete example, which motivated this paper's work,
is when the observable data \vecc\ are the binned 2-point correlation functions of
cosmic fields derived from a catalog of galaxies; the parameters of
interest are cosmological quantities such as the  matter density
$\Omega_m,$ the amplitude of density fluctuations $\sigma_8,$
etc.; and the nuisance parameters $\vecn$
include the coefficients of some linear expansion of the redshift distributions $n(z)$ of the galaxies being observed:
\begin{equation}
  n(z) = \sum_{k=1}^{N} n_k b_k(z).
  \label{eq:nzbasis}
\end{equation}
The $b_k$ are a set of predetermined basis functions for the redshift
distribution, \eg\ these are boxcar functions if we are modeling $n(z)$ as
  stepwise constant.  In our case of analyzing the data from the  \textit{Dark Energy Survey} (DES), there are 10 distinct samples of galaxies---each designed to prefer galaxies in a limited range of redshift---which we can index by $s.$ Each has its own $n_s(z)$ to be characterized by coefficients $n_{sk}$ at $\approx100$ values of $k$, leading to $N=O(1000)$ parameters $n_{sk}$ to be considered.  The vector $\vecn$ of nuisance parameters would be the concatenation of all the $n_{sk}.$ For clarity, we will still write this as $\vecn=\{n_1,n_2,\ldots,n_N\}$ and demonstrate the method with a single galaxy sample's $n(z)$.

One approach would be to run a new MC over \vecq\ for each of the samples
we have of \vecn, and then concatenate these to effect marginalization over
\vecn.  This is clearly infeasible if thousands or more of \vecn\ samples are needed to characterize the prior in this space.

Facing this problem for the cosmological analyses of the 3-year data
(Y3) from DES,
\citet{hyperrank} devised a scheme whereby
the samples of \vecn, which we write as $\{\vecn_\alpha\}$ for $\alpha\in
  1\ldots N_{\rm samp},$ are assigned to equally-spaced grid points within some $M$-dimensional 
unit hypercube $\mathcal{H}$.  The coordinates \vecu\ within the hypercube
are considered the compressed parameters of $n(z),$ and the
decompression function
$\hat{\vecn}(\vecu)$ outputs the $\vecn_\alpha$ sample at the nearest grid point to
any  \vecu, \ie\ nearest-neighbor interpolation.
In the example
  application described in Section~\ref{sec:app}, this procedure would assign an
  $N=80$-dimensional \vecn\ to each grid point placed in an $M=3$-dimensional hypercube.
This solves the problem of creating a continuous \vecu\
  domain, and gives each sample $\vecn_\alpha$ equal probability under
  $p(\vecu),$ maintaining the meaning of the input samples.
But the \emph{output} of the function $\hat\vecn(\vecu)$, and the resultant likelihood
  function $\likeli(\vecc | \vecq, \vecu),$ are discontinuous over $\vecu.$
Various strategies are proposed by \citet{hyperrank} to assign the $\vecn_\alpha$ to the grid points in
$\mathcal{H}$ based on summary statistics, to reduce the
discontinuities---but the function is never smooth.
As a consequence, many MC samplers become quite inefficient in
sampling of the cosmological posterior.  In particular, samplers such
as \textsc{MultiNest} that assume continuity are rendered nearly
non-functional.  As a result, the Y3 cosmological priors could not be
evaluated with this method.  Instead, the \vecn\ samples were not
used, and an \textit{ad hoc} $p(\vecn)$ was adopted which allowed only
shifts and dilations of the mean $n(z)$ of the \vecn\ samples [see \eqq{eq:zs}].

A more rigorous and extremely efficient method of marginalizing over high-dimensional nuisance parameters was 
described by \citet{bridle02} and reprised by \citet{hans} for the $n(z)$ application, for the case where the following restrictions apply:
\begin{enumerate}
\item The likelihood of the observable \vecc\ is normal, $\vecc \sim \mathcal{N}( \hatc, \covm_c),$ with $\covm_c$ fixed.
\item The prior $p(\vecn)$ can also be assumed to be normal, with a mean taken to be $\bar\vecn = \left\langle\vecn\right\rangle$ and covariance matrix taken to be $\covm_n=\left\langle(\vecn-\bar\vecn)(\vecn-\bar\vecn)^T\right\rangle$ using the samples of \vecn\ we are given.
\item The model \hatc\ can be linearized in \vecn\ about fiducial values $\vecq_0, \vecn_0$ without loss of accuracy exceeding measurement errors, with the derivatives independent of \vecq.
\end{enumerate}
Under these conditions, the marginalization over \vecn\ is shown to be
  algebraically equivalent to adding terms to $\covm_c,$ such that any MC process need not sample \vecn\ at all.  

We describe here an approach that is algebraically similar to this analytic marginalization,
but does not require the 2nd condition of Gaussianity for
the nuisance prior.
Our approach is to seek a linear compression of \vecn\ into a lower-dimensional
set of parameters \vecu\ that projects away variations in \vecn\ that do not
influence the likelihood $\likeli.$  Standard density estimators can then be
applied to the \vecu\ values implied by the known \vecn\ samples to yield a
prior $p(\vecu)$ that can be used for the MC of the cosmological posterior.  The
model \hatc, and hence $\likeli,$ will be continuous over this low-dimensional
\vecu\ space, and marginalization over \vecu\ will yield posterior probabilities
very close to marginalization over the original \vecn.  

\section{Derivation}\label{sec:deriv}
We assume that we do have a multivariate normal likelihood for the
observables \vecc\, with the mean being some model
$\hatc(\vecq,\vecn)$ and a fixed covariance matrix $\covm_c,$ known \textit{a
  priori} via analytic calculations and/or jacknifing of the data or other methods.   In this
section we will assume that the \vecn\ vectors have been shifted by
the mean of the samples $\vecn_0 \equiv \left\langle \vecn_\alpha
\right\rangle,$ generating a replacement set of $\vecn_\alpha$ that have zero mean.
We seek some function $\hat{\vecn}(\vecu)$ of a much lower-dimensional
vector \vecu\ which can be substituted for \vecn\ and yield nearly the
same likelihood function for any \vecn\ in the domain spanned by the
samples $\{\vecn_\alpha\}.$
This means we want maps $\vecn_\alpha\rightarrow
\vecu_\alpha \rightarrow \hat{\vecn}_\alpha,$ with ${\rm dim}(\vecn)={\rm dim}(\hat{\vecn})=N$ and ${\rm dim}({\vecu})=M\ll N.$ We wish to find maps such that replacing $\vecn$ with
$\hat\vecn$ alters the cosmological inference by much less than the
other uncertainties in the model or data.
Inferences use the likelihood of the observed $\vecc$ under the
assumed Gaussian probability distribution:
\begin{equation}
  -2\log\likeli(\vecc | \vecq,\vecn) = |2\pi C_c| + \left[\vecc-\hatc(\vecq,\vecn)\right]^T
  C^{-1}_c\left [\vecc-\hatc(\vecq,\vecn)\right].
\label{eq:like1}
\end{equation}
While ideally the loss function for a compression scheme would be some direct
measure of the biases induced in the inferred parameters \vecq, we will instead
target a minimization of the expected level of mis-estimation of the likelihood
function $\likeli.$ The posterior credible region for \vecq\ at fixed
\vecc\ will correspond to a range of $\Delta(2\log\likeli)\approx 1$ (unless
dominated by priors.)  Thus we can be confident that a compression that changes
the inferred $\log\likeli$ by $|\Delta\log\likeli|\ll 1$ will change
inferred \vecq\ values by a small fraction of the measurement uncertainty.
It is certainly true that as $\langle
\Delta\log\likeli \rangle \rightarrow 0,$ the substitution of $\hat\vecn$ for $\vecn$
has no effect on the calculated posterior density of \vecq.
We
do not, however, offer any formal proof or bound on the relation between
$\log\likeli$ errors and \vecq\ errors induced by compression.

If the compression scheme replaces $\vecn$ with $\hat\vecn,$ then we will assign
an incorrect $\hat\likeli$ 
\begin{align}
\label{eq:like2}
   -2\log\hat\likeli & = |2\pi C_c| + \left[\vecc-\hatc(\vecc,\vecn)-\Delta_c\right]^T
                      C^{-1}_c\left [\vecc-\hatc(\vecq,\vecn) - \Delta_c\right], \\
  \Delta_c & \equiv \hatc(\vecq,\hat\vecn) -  \hatc(\vecq,\vecn).
\end{align}
Expanding (\ref{eq:like2}) and subtracting (\ref{eq:like1}) yields the
compression error in the likelihood:
\begin{equation}
 \Delta(-2\log\likeli) =  \Delta_c^T C^{-1}_c \Delta c +
                        2\left[\vecc-\hatc(\vecq,\vecn)\right]^T C_c^{-1} \Delta_c.
\label{eq:like3}
\end{equation}
If the model $\hatc(\vecq,\vecn)$ properly describes the mean of the
observations $\vecc,$ then the expectation value of the rightmost term under the
observational noise is zero. A useful measure of the typical mis-estimation of
$\vecq$ induced by
the compression of \vecn\ is therefore the term $\Delta_c^T C^{-1}_c \Delta_c.$
Marginalizing this quantity over the (sampled) prior for $\vecn$ yields the loss
function that we will minimize during compression:
\begin{equation}
 \left\langle \chi^2 \right\rangle =  \frac{1}{N_{\rm samp}} \sum_\alpha
                                            \left[ \hatc(\vecq,\vecn_\alpha) - \hatc(\vecq,\hat{\vecn}_\alpha) \right]^T
                                            \covm_c^{-1}
                                            \left[ \hatc(\vecq,\vecn_\alpha) - \hatc(\vecq,\hat{\vecn}_\alpha) \right].
\label{eq:chisq}
\end{equation}
This quantity is the $\chi^2$ of the difference
between the original and compressed models for the data, marginalized over the
prior on \vecn\ as represented by the samples we are provided.
It is also the mean-squared distance
in the space $\vecc$ between the
model generated by $\vecn$ and that by $\hat\vecn$, using the
observations' covariance matrix $\covm_c$ as a metric for the
distance.\footnote{More precisely, the inverse of the covariance
  matrix is the metric.}

The use of \eqq{eq:chisq} implicitly assumes that $N_{\rm samp}$ is sufficiently
large that averaging $\chi^2$ over the samples is equivalent to integrating over the
entire $\vecn$ space.    One might be concerned that a compression scheme could
overtrain on \eqq{eq:chisq} in the sense of yielding small $\chi^2$ at the
sample points $\vecn_\alpha,$ but have large $\chi^2$ at \vecn\ values between
the samples.  If, however, we use a simple linear  projection for compression to $M$
dimensions, there are $NM$ free parameters, whereas the input samples have
$NN_{\rm samp}$ values to operate on.  If $N_{\rm samp}\gg M,$ we assert without
proof that the compression
function does not have enough freedom to overtrain except in pathological cases
of sampling.  
But our scenario already assumes that the samples are
sufficient to conduct marginalization over $\vecn$ when estimating
$p(\vecc|\vecq,\vecn),$ which means they must be numerous $(N_{\rm samp}\gg N)$ and span the space of
plausible $\vecn$ values under $p(\vecn).$ 

We next assume that the compression is linear, $\hat\vecn=\matX\vecn,$
for some $N\times N$ matrix that is idempotent ($\matX\matX = \matX$).  We will further linearize the dependence of $\hatc$ on $\vecn,$ specifically assuming that (in scalar notation) 
\begin{equation}
    \frac{\partial^2 \bar c}{\partial n^2} n \ll \frac{\partial \bar c}{\partial n}
\label{eq:linearmodel}
\end{equation}
over the full range of variation of $\vecn.$  
With these two assumptions, \eqq{eq:chisq} becomes
\begin{equation}
  \left\langle \chi^2 \right\rangle = \frac{1}{N_{\rm samp}} \sum_\alpha
  \left[ (\matI-\matX)\vecn_\alpha\right]^T \matF  \left[ (\matI-\matX)\vecn_\alpha\right].
  \label{eq:linearized}
\end{equation}
We use the Jacobian matrix of the model $\hatc$ to define
\begin{equation}
  \matF \equiv
  \left[\frac{\partial\hatc}{\partial\vecn}\right]_{\vecq_0, \vecn_0}^T
  \covm_c^{-1} \left[\frac{\partial\hatc}{\partial\vecn}\right]_{\vecq_0,
    \vecn_0}.
\label{eq:fisher}
\end{equation}
This quantity is also the Fisher matrix giving the information
provided by the observations $\vecc$ about the nuisance parameters
\vecn\ under a Gaussian likelihood \citep{TTH}.  In many cases this matrix will be
rank-deficient and/or poorly conditioned, since the observables are not
likely to be very informative on $\vecn$---if they were, we might not be
concerned with establishing a prior on $\vecn$ to begin with.
Fortunately, we will not require the inverse of $\matF$ in our algorithm.

Our assumptions that the likelihood is normal in $\hatc$ and that we can
  linearize $\hatc$ in $\vecn$ [equations (\ref{eq:like1}) and
  (\ref{eq:linearmodel})] can be replaced by the slightly more general condition
  that the Hessian of $\log\likeli$ with respect to $\vecn$ is well approximated
  as constant over the domain of interest of $(\vecq,\vecn)$.

The optimization implied by \eqq{eq:linearized} is the same as in
familiar Principal Components Analysis (PCA), aside from the presence
of the $F$ matrix, which in essence defines a new metric for the
variance to be captured by the principal components.  Our solution will follow the typical derivation
for PCA, but with an additional variable transformation to compensate
for the presence of $\matF.$

Since $\matX$ is idempotent, we can write
\begin{equation}
  \matX = V_X \proj_M V_X^T,
\end{equation}
where $V_X$ is unitary and the projection matrix $\proj_M$ is defined as
\begin{equation}
  \left(\proj_M\right)_{ij} \equiv
\begin{cases}
                                            1,  &  i=j\le M \\
                                            0,  & \text{otherwise}
\end{cases}
\end{equation}
It is also useful to define
\begin{equation}
  \matY \equiv \matI-\matX  = V_X \proj_{-M} V_X^T,
\end{equation}
with $\proj_{-M}=\matI-\proj_M.$  

For a chosen rank $M$ of the
transformation matrix $X$, our task becomes to identify the
eigenvectors $V_X$ that minimize
\begin{align}
  \left\langle \chi^2\right\rangle & = \frac{1}{N_{\rm samp}} \sum_\alpha
  \left[ \matY \vecn_\alpha\right]^T \matF  \left[\matY
                                     \vecn_\alpha\right] \\
       & = \trace \left[ \covm_n V_X \proj_{-M} V_X^T \matF V_X
         \proj_{-M} V_X^T \right], \\
  C_n & \equiv \left\langle \vecn \vecn^T \right\rangle.
\end{align}
This optimization is easier if we first transform the systematic variables to
$\vecn^\prime =\matT\vecn$ such that $\covm_{n^\prime}=\matI$, \ie\
make the elements of $\vecn^\prime$ uncorrelated and unit-variance.  This
is accomplished by finding the eigensystem $\covm_n=\matV_n \matLam_n
\matV_n^T$ and setting $\matT = \matLam_n^{-1/2} \matV_n^T$.  With
this transformation, we are now seeking a different unitary matrix
$\matV_{X^\prime}$ that minimizes
\begin{align}
  \left\langle \chi^2\right\rangle & = \trace \left[ \matI
    V_{X^\prime} \proj_{-M} V_{X^\prime}^T \left[ (T^{-1})^T \matF
      T^{-1} \right] V_{X^\prime} 
    \proj_{-M} V_{X^\prime}^T \right] \\
  & = \trace \left[ \proj_{-M} V_{X^\prime}^T \matG 
    V_{X^\prime} \proj_{-M} \right],
  \label{eq:mintr}
\end{align}
where we have defined the transformed Fisher matrix
\begin{equation}
  \matG \equiv \left(T^{-1}\right)^T \matF   T^{-1} = \matLam_n^{1/2} \matV_n^T
  \matF \matV_n \matLam_n^{1/2} = \matV_G \matLam_G \matV_G^T.
  \label{eq:defG}
\end{equation}
The right-hand side defines the eigensystem of $G$, with $\Lambda_G={\rm diag}(\lambda^G_1,\ldots,\lambda^G_N),$ and $\lambda^G_i\ge0.$  \eqq{eq:mintr} can now be rewritten as
\begin{equation}
\left\langle \chi^2\right\rangle = \sum_{i>M} \left(\matV\lambda_G \matV^T\right)_{ii}
\end{equation}
where $\matV=\matV^T_{X^\prime}\matV_G$ is unitary.  This expression must be at least as large as sum of the $N-M$ smallest $\lambda^G_i$, and that minimum is attained if
$\matV_{X^\prime}^T \matV_G = \matI \; \Rightarrow \; \matV_{X^\prime} = \matV_G,$ and the eigensystem of $G$ is placed
in order of decreasing eigenvalues $\lambda^G_i.$ The
elements surviving the projection $P_{-M}$ yield
\begin{equation}
  \left\langle \chi^2\right\rangle = \sum_{i>M} \lambda_i^G.
  \label{eq:chiresid}
\end{equation}
In other words each eigenvalue of the matrix $\matG$ in \eqq{eq:defG}
gives the contribution to $\left\langle\chi^2\right\rangle$ of one
projection (mode) of $\vecn.$

Transforming the solution back into the space of $\vecn$ yields
\begin{align}
  \matX & = \matT^{-1} \matV_G \proj_M \matV_G^T \matT \\
\label{eq:DE}
   & = \left[ \matV_n^T \matLam_n^{1/2} \matV_G \proj_M \right] \left[
     \proj_M \matV_G^T \matLam_n^{-1/2} \matV_n^T \right] \\
        & \equiv \matD \matE.
\end{align}
We thus obtain our optimal encoding/compression using the nonzero rows of matrix
$\matE$ to give
\begin{equation}
  \vecu_\alpha = \matE \vecn_\alpha
  \label{eq:uu}
\end{equation}
and the decoding/reconstruction of the systematic variables as
\begin{equation}
  \hat\vecn_\alpha = \matD \vecu_\alpha.
  \label{eq:UU}
\end{equation}
One can confirm that this procedure yields a compressed representation
$\vecu$ such that $\covm_u = \matI_M,$ the $M$-dimensional
identity matrix.

The previous derivation ignores the possibility that $\covm_n$ is
singular or nearly so, such that taking $\matLam_n^{-1/2}$ in
\eqq{eq:DE} is not possible.  Indeed in our application, it is
\emph{required} that $\covm_n$ be singular, because we have a sum
normalization constraint on the initial $\vecn_\alpha$ values.  Any
such (nearly) zero element $j$ of $\matLam_n$ has a corresponding
eigenvector $\vecv_j$ of the $\vecn$ space which has zero amplitude
in all of the input samples $\vecn_\alpha,$ so that the $\vecn$ are
confined to a subspace---therefore the reconstructed $\hat\vecn$ should also be.
The compressed
representations $\vecu_\alpha$ and reconstructed $\hat\vecn_\alpha$
should therefore be unaffected by the presence of any $\vecv_j$
component.  This can be accomplished in \eqq{eq:DE} by setting element
$j$ of $\matLam_n^{-1/2}$ to zero, as is typically done during
solutions of least-squares problems using singular value
decompositions.

In summary, the procedure for dimensional reduction is:
\begin{enumerate}
  \item Obtain the mean $\vecn_0$ of the input samples, the Fisher matrix $\matF$ of the system defined in
    \eqq{eq:fisher} using derivatives about $\vecn_0,$ 
    plus the covariance matrix $\covm_n$ of the samples.
  \item From the eigensystem $(\matLam_n,\matV_n)$ of $C_n$, form the
    matrix $G$ defined in \eqq{eq:defG} and get its eigensystem
    $(\matLam_G,\matV_G).$  Place the eigenvalues in descending order.
  \item Choose the size $M$ of the compressed representation to be the
    minimum that keeps the $\left\langle\chi^2\right\rangle$ value in
    \eqq{eq:chiresid} below a chosen threshold, presumably $\ll 1.$ If this
      is achieved, the multiplicative errors in the likelihood induced by the
      compression are at percent
      levels. 
  \item Form the encoding matrix $E$ and decoding matrix $D$ as
    in \eqq{eq:DE}, taking the inverse $\matLam_n^{-1/2}$ to be zero
    for any eigenvalues that are zero (or within roundoff errors).
  \item Compress all incoming (mean-subtracted) samples using
    \eqq{eq:uu}.  The resulting $\vecu$ values will have
    have unit covariance matrix and zero mean.
  \item Construct a continuous density estimator in $\vecu$ space that
    mimics the finite sample distribution.  If the distribution is
    normal, this becomes the multidimensional unit normal, since the
    construction yields $\textrm{Cov}(u)=I_M.$.  There
    is, however,  no \textit{a priori} reason that this must be the case,
    and something like a normalizing flow may be needed to
    approximate this lower-dimensional representation of the prior.
  \item Sample over $\vecu$ space in the Markov Chain that is sampling the
    posterior on the parameters of interest $\vecq,$ using \eqq{eq:UU}
    to transform each sample back into a $\hat\vecn$ vector.
  \end{enumerate}
  
The ability to accomodate non-Gaussian distributions of the
nuisance-parameter space is the principle
advantage of our method over the single-step covariance-inflation
method of \citet{bridle02} and \citet{hans}, \ie\ one can drop assumption (2) of the three listed for
this method in the Introduction.
 But the compression method is also more robust in other ways: it defines
   projection of \vecn\ values onto an $M$-dimensional linear subspace.  We can
   in fact also drop assumptions (1) of Gaussian likelihood and (3) of linearization
   of the model $\hatc(\vecq,\vecn)$ that were used to derive the subspace,
   if we can instead demonstrate that: \emph{no component of \vecn\
     normal to the subspace can significantly alter the likelihood of the data,
     when $\vecn$ and $\vecq$ range across their regions of significant
     probability.}  If this statement is true, then marginalization over the
   compressed subspace will yield inferences consistent with the (usually
   infeasible) marginalization over the full space of \vecn, regardless of the
   nature of the model and likelihood functions.  

%
Even in the case where all three assumptions of the covariance-inflation
method hold true, 
there are practical advantages of compressing the nuisance 
variables and retaining them in the cosmological Markov chain instead of doing
the analytic marginalization.   One can, for example, examine the posterior
distribution of \vecu\ to see how the data have altered the prior
  distribution of $\vecu.$  If the posterior for $\vecu$ is at the
edge of the prior, this would potentially indicate an inconsistency between the
data and the prior.

\section{Application}\label{sec:app}
As an example of the application of this straightforward dimensional
reduction to a high-dimensional nuisance parameter, we examine the
redshift distribution of one of the bins of ``Maglim''  galaxies used
as a lens population and clustering tracer in the Year 6 (Y6) analysis of
the DES galaxy catalogs.  Each of these Maglim ``lens bins''
is selected with cuts on galaxy fluxes and
colors in an attempt to generate a sample that is confined to a
particular redshift range, as described in \citet{y6maglim}.  Once each bin's
galaxy selection criteria are chosen, a combination of 
photometric techniques \citep{y6lenspz,y6pz} and clustering information
\citep{y6wz} is used to generate samples from the posterior
distribution of  $n(z)$ functions applicable to the bin members, conditioned
on the photometric and clustering data, \ie\ the $p(\vecn)$ that will become the
  prior for the inference of \vecq.

In the simplest case,
there are 6 cosmological parameters of interest, $\vecq = \{\Omega_m,
\Omega_b, \sigma_8, h, n_s, m_\nu\}.$   The nuisance vector $\vecn$
has $\approx 5$  parameters in each of the following categories: galaxy
biases with respect to matter; intrinsic alignments of galaxy shapes
with the tidal field of the mass; magnification coefficients; and multiplicative errors in the
measurement of galaxy shear, for a total of 24 non-redshift parameters.  The $n(z)$ for each galaxy bin
is
described by \eqq{eq:nzbasis} with 80 coefficients spanning $0<z<4$ at
intervals of $\Delta z=0.05.$ If all of these coefficients, for each
of the 10 galaxy selections, were allowed to vary, the parameter space for inference would grow from 24 to 824 dimensions.
The inference would become infeasible even if a
reliable density estimator over the 80-dimensional space could be
created from the $O(10^4)$ samples available to characterize each
$n(z).$ 

We hence turn to the modal projection technique herein to reduce the
dimensionality to those directions in $\vecn$ space in which the samples span a large
enough range to alter the $\hatc(\vecn)$ at levels comparable to the
  precision of the observations.

The observable quantities are the autocorrelation $w(\theta)$ of the 
galaxies' angular positions as a function of angular separation $\theta$;
  and the cross-correlations $\gamma_t^{(s)}(\theta)$ between 
these galaxies' positions and the weak gravitational lensing shear
observed from groups $1\le s \le 4$ of source galaxies.  The vector $\vecc$ is
the concatenation of values of $w$ and $\gamma_t$ in a set of bins of $\theta.$
The model $\hatc(\vecq,\vecn)$ for the observables consists of integrations
  over redshift $z$ of products of the redshift distributions $n(z)$ with
  solutions of General Relativity differential equations for the expansion
  history and growth of density fluctuations in the Universe.  The covariance
  matrices $\covm_c$ are even more complex.  The derivations of both are
  described in full detail in 
 \citet{y6model}. We use the same \texttt{Cosmosis} software \citep{cosmosis}
 employed in that paper to
calculate the partial derivatives needed in deriving the compression scheme.

\begin{figure}
  \center
  \includegraphics[width=0.6\textwidth]{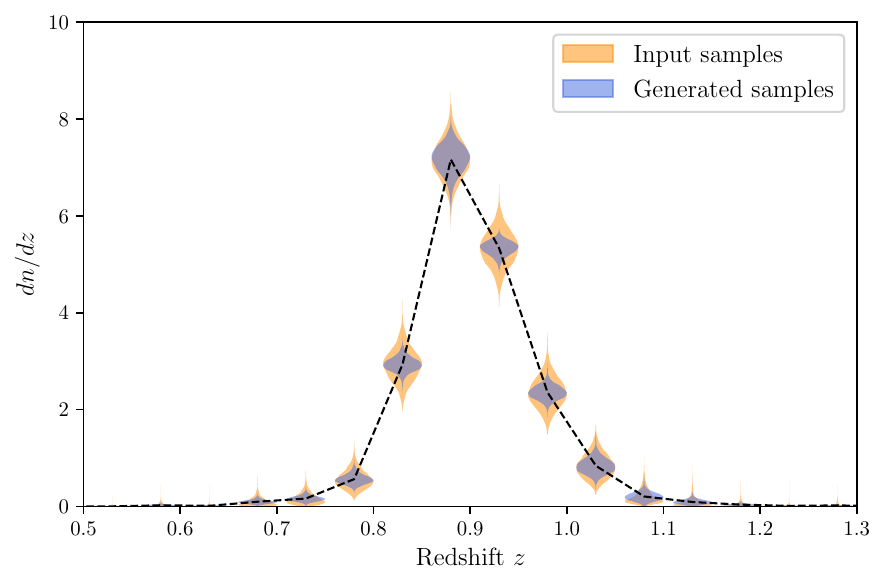}
\caption{Violin plots for the redshift probability distribution $n(z)$
  of galaxies in lens bin 4 for the DES Y6 analysis.  The orange regions
  show the distributions for the samples of $\vecn$ derived from
  photometric and clustering information.  The blue violins are for
  $\vecn$ values drawn defined by (1) subtracting the mean $\bar\vecn$; (2) compressing these $\vecn$
  into 3 modes with coefficients $\vecu;$ (2) drawing values of
  $\widetilde{u}_i$ from unit normal distributions; (3) transforming each
  component $\widetilde u_i$ to match the 1d distribution of the input samples' $u_i$; (4) decompressing the transformed $u_i$ draws back into full-length $\vecn$ samples; finally (5) restoring the mean $\bar\vecn.$  The dashed
  line connects the mean values of $n(z),$ which are the same for
  generated samples as for the input samples, by construction.  The compression substantially lowers the variance of $n(z)$ at individual values of $z$ without significantly altering the variation of cosmological signals that the entire $n(z)$ predicts.
  [Although the $n(z)$ functions are calculated for $0<z<4,$ we
  truncate this plot to emphasize the
  redshift regime where this bin's galaxies are primarily found.]}
  \label{fig:violins}
\end{figure}

We present results for mode-compression sampling of the $n(z)$
parameters for redshift bin 4 of the DES Y6 lens galaxies.  The orange violins
in Figure~\ref{fig:violins}
plot the distributions of the individual elements of $n(z)$ in the
input 3000 samples.  As per the procedure described in this paper, the
mean \vecn\ is subtracted from each sample and the covariance
$\covm_n$ computed.  This is then combined with the
derivatives and covariance matrices of the observables $\vecc$ to
calculate the decomposition in \eqq{eq:defG}.

\begin{figure}
  \center
  \includegraphics[width=0.49\textwidth]{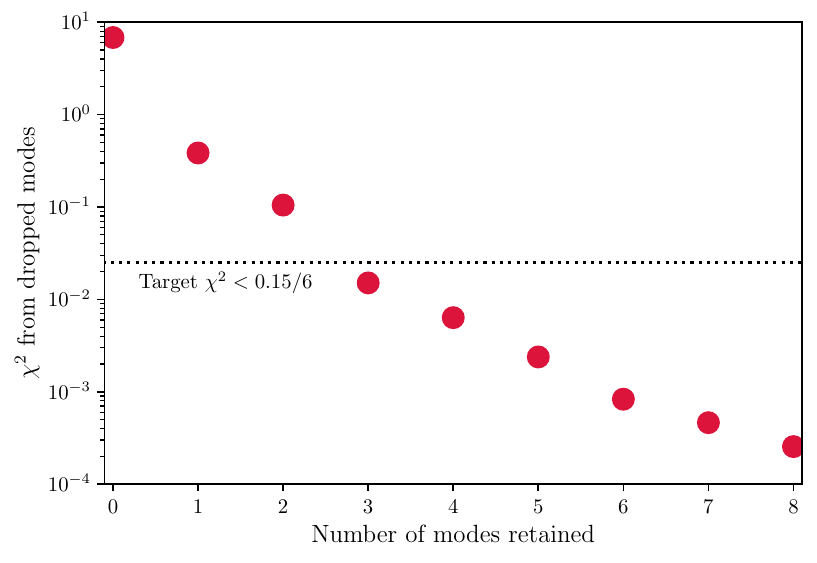}
   \includegraphics[width=0.49\textwidth]{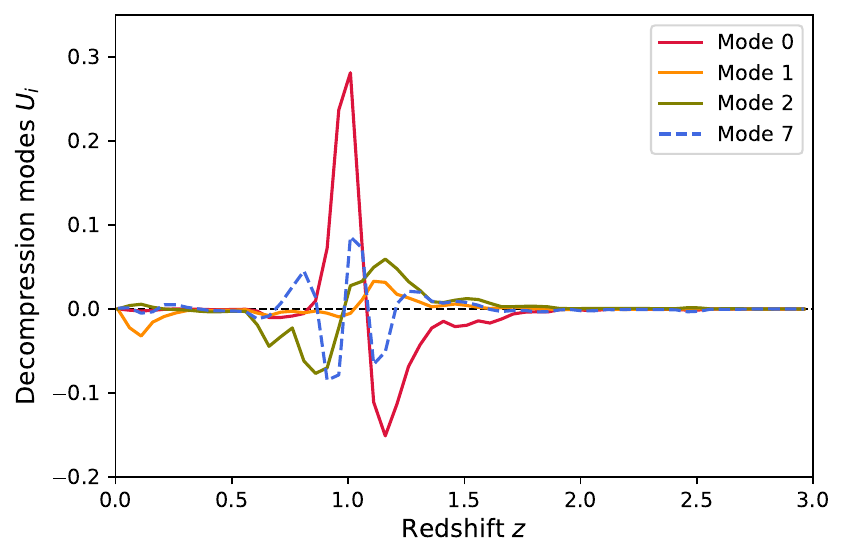}
\caption{At left: The size of the $\chi^2$ of modeling error attributable to compressing the
  $\vecn$ samples down to $M$ modes is plotted vs $M$.  The $M=0$ point shows the modelling error from holding $n(z)$ fixed at its mean, and $M\ge1$ values drop exponentially as we use more modes to reconstruct $n(z).$ Our chosen criterion of
  $\chi^2<0.025$ is attained with $M=3$ for this bin's
  $n(z).$
  At right: The modes of variation $U_i(z),$ \ie\ the rows of the decompression
  matrix $\matD$ in \eqq{eq:DE}, are plotted vs redshift.  Each of
  modes 0,1,2 is
  multiplied by a unit-variance stochastic coefficient $u_i$, then
  they are summed with the mean $\bar\vecn(z),$ to form an $n(z)$
  sample.  Higher-numbered modes have
  observable consequences of decreasing statistical significance. Mode 7 is plotted as the dashed blue line as an example of what is projected out of $n(z)$; even though its typical amplitude in the input data is larger than modes 1 or 2, its oscillatory behavior does not lead to measurable changes in the cosmological statistics.}
  \label{fig:chiresid}\end{figure}

The left plot in Figure~\ref{fig:chiresid} shows the values of unmodelled
$\chi^2$ vs the dimension $M$ of the compressed space, as per
\eqq{eq:chiresid}.  This modelling error induced by compression drops
exponentially with the number of retained modes.  We have somewhat arbitrarily
chosen a threshold of $\chi^2<0.025$ for each galaxy sample in order to keep the
total impact of compression of 10 samples to $\ll 1.$ This is attained with $M=3$ modes for this bin.  The right side of Figure~\ref{fig:chiresid} plots the individual modes
${\rm U}_i(z),$ \ie\ the rows of the decompression matrix $\matD$ such
that $\hat\vecn = \bar\vecn + \sum_{i\le M} u_i \vecU_i.$ Recall that
each mode's coefficient $u_i$ will be a random deviate with unit variance. Each
of the first three modes appears to effect some combination of a $z$ shift
of the main $n(z)$ peak, a change of the peak's shape/width, and a change in the
low-$z$ contamination. We have also plotted mode 7 in the Figure, to illustrate
a mode of variation that is present in the samples, but has unobservable
consequences.  Mode 7 is more oscillatory in redshift than the three retained
modes, and does not alter the low-$z$ tail. 

\begin{figure}
  \center
  \includegraphics[width=0.7\textwidth]{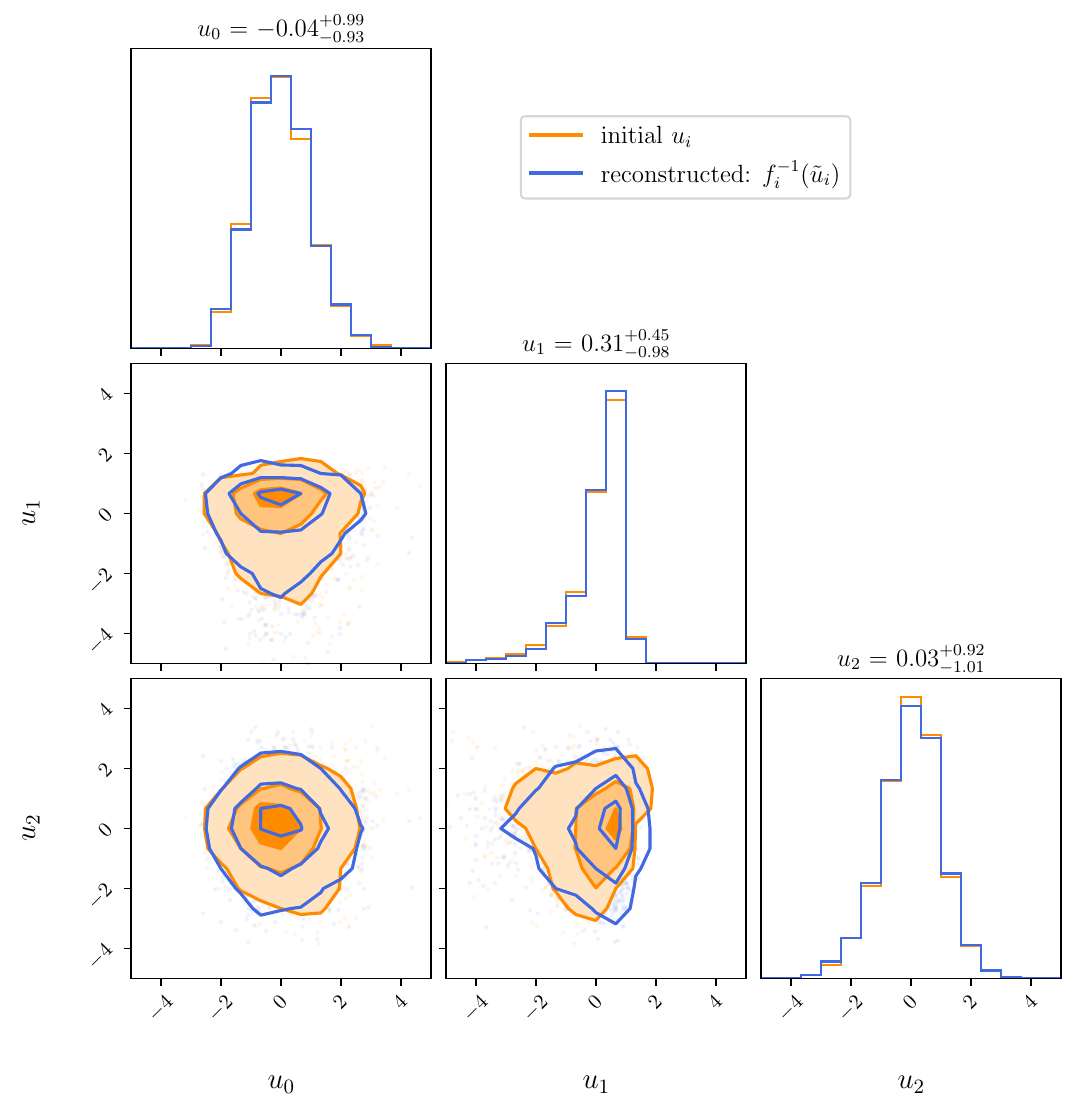}
  \caption{In orange is a corner plot of the distribution of the mode coefficients,
    \ie\ the elements of $\vecu = \matE \vecn.$
  of the input samples after encoding.  The coefficients, especially 
  $u_1,$ are significantly skewed so a normal distribution would be an
  inaccurate model.  Instead we model each $u_i$ as a
  ``denormalizing'' function of a unit-normal variable, as per
  \eqq{eq:denorm}.  The blue histograms and contours show the
  distributions obtained using this element-by-element transformation
  technique, which is seen to accurately reproduce the distribution of
  the input samples.}
\label{fig:corner}
\end{figure}

Figure~\ref{fig:corner} plots the distributions
of the $M=3$ components of the compressed $\vecu$ representations of the
3000 input samples, \ie\ the vectors $\vecu_j=E(\vecn_j-\bar\vecn)$ obtained from each input sample $\vecn_j$.  Some of the components of $\vecu$ have substantially non-Gaussian distributions,
so a unit-normal prior on $\vecu$ will not faithfully represent the
input samples---a better density estimator is required.  We find in
this case (and in all other DES cases) that it is sufficient to
normalize the marginal distributions of the individual $u_i$
components.  This is done by tabulating a normalizing function $f_i$
for each mode defined by
\begin{equation}
  {\rm CDF}_n\left[ f_i(u_i) \right] = {\rm CDF}(u_i),
\label{eq:denorm}
\end{equation}
where the left side is the cumulative distribution function of the
unit normal, and the right-hand side is the CDF of the $u_i$ values
obtained from the input samples.  The functions $f_i$ are bijective
and can be stored as a splined lookup table.
Now, the cosmological Markov Chain
is told that there are 3 parameters in $\widetilde{\vecu}$ that have a
unit-normal prior $p(\widetilde{\vecu}).$  This vector defines the
redshift distribution via a 2-step process:
\begin{align}
  u_i & =f_i^{-1}(\widetilde{u}_i); \\
  \vecn & = \matD \vecu.
\end{align}

The blue violins in Figure~\ref{fig:violins} show the distributions
of the $n(z)$ values that are generated by drawing $\widetilde{\vecu}$
values from a unit Gaussian.
The means of the two distributions (centers of the violins)
  are, by construction, identical.  The blue, compression-derived distributions
  are shorter (narrower distributions) than the orange input distributions,
  which is a result of projecting away modes of \vecn\ that have no measurable influence on the
  observable \vecc\ values.  Removing modes reduces the variance of $n(z)$ at any
  particular $z$, by varying degrees at different values of $z.$ The compression
scheme acts as noise reduction on the $n(z)$ estimates, where ``noise'' means a
fluctuation that does not detectably impact measurable quantities.
Projecting away the
unobservable fluctuations in $n(z)$ has substantially reduced the
variance of the function at any \emph{individual} $z$ value, but the
\emph{collective} $n(z)$ behavior still retains the same observable influence on
the summary statistics $\vecc$.

To check whether we have achieved our goal
of leaving the observable consequences of the $n(z)$ variation
unchanged, we calculate the distribution of
\begin{equation}
  \chi^2 =  \left[\hatc(\vecq,\vecn) - \hatc(\vecq,\vecn_0) \right]^T
                                           \covm_c^{-1}  \left[ \hatc(\vecq,\vecn) - \hatc(\vecq,\vecn_0)\right]
\label{eq:chihist}
\end{equation}
for the cases when (1) the $\vecn$ are drawn from the input samples, vs
(2) are generated using the procedure defined above.  This $\chi^2$
measures the deviation of the model from that implied by the mean
vector $\vecn_0$ at some chosen nominal value of cosmological
parameters $\vecq$ (we also fix the other nuisance parameters of the
DES model for this test).\footnote{Please note that this $\chi^2$ quantity
    is generated from nuisance parameter samples, not by any process that should
    reproduce the standard $\chi^2$ distribution.}

\begin{figure}
  \center
  \includegraphics[width=0.6\textwidth]{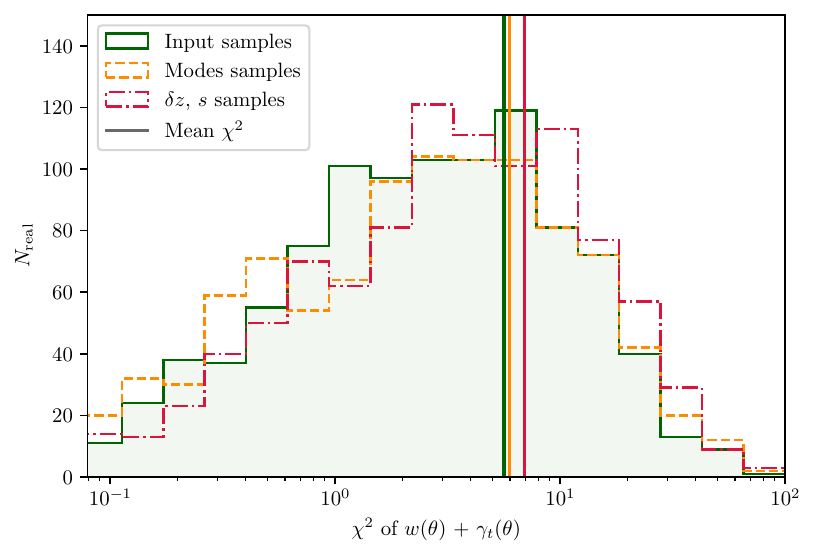}
\caption{The histograms show the deviations of the predicted observable $w(\theta)$ quantities in
  DES Y6 cosmological analysis, as measured by the $\chi^2$ in
  \eqq{eq:chihist}, as we allow the $n(z)$ parameters to vary.  The
  shaded green histogram shows the variation using the original 3000 samples of
  $n(z)$ produced by the photometric and clustering redshift studies.
  The dashed yellow histogram results from
  drawing 3-dimensional $\widetilde{\vecu}$ values from a unit
  normal and decompressing them into $n(z)$ realizations using our
  method.  This lower-dimensional model for the prior on $n(z)$
  reproduces the original samples' result very well.  By comparison,
  the dash-dotted red histogram generates samples of $n(z)$ using
  the \textit{ad hoc} method of \eqq{eq:zs}.  This method's two
  parameters $z$ and $s$ are given priors to match the distributions
  of the mean of $n(z)$ over $z$, and the standard deviation of $z$,
  present in the input samples.  The \textit{ad hoc} method produces
  $n(z)$ fluctuations with $\approx20\%$ larger $\chi^2$ from the mean,
  on average, than the input distribution has. [Note the logarithmic horizontal
    axis.]} 
\label{fig:chihist}
\end{figure}

Figure~\ref{fig:chihist} shows the results: the distribution of the $\chi^2$
  values defined by \eqq{eq:chihist}
using
the input samples is indistinguishable from the distribution of this $\chi^2$
using samples of \vecn
generated by our compressed, normalized representation.
Note that this agreement indicates that the approximation of linearity of the
  model in \eqq{eq:linearmodel} is sufficiently accurate in this application to
  yield an accurate compression scheme.
The Figure also shows the $\chi^2$ distribution resulting from samples
generated by the method used in the DES Y3 analysis \citep{y3pz}.  In
that case, the $n(z)$ for each galaxy sample was given an \textit{ad
  hoc} variation of the form
\begin{equation}
  n(z) = n_0\left(\frac{z-\Delta\,z}{s}\right),
  \label{eq:zs}
\end{equation}
with $\Delta z$ and $s$ being ``shift'' and ``stretch'' parameters.
Separable Gaussian priors were assigned to $\Delta z$ and $s$ with
means of 0 and 1, respectively, and standard deviations that equaled
the RMS variation in the mean and width of the input $n(z)$ samples.
The shift-stretch model essentially compresses the samples of $n(z)$
into two parameters, their mean and standard deviation, and executes
an \textit{ad hoc} reconstruction based on those parameters.
In the Figure we can see that the shift-stretch model does not in fact
reproduce the size of the deviations in the $\vecc$ observables that
is implied by the original samples; the mean $\chi^2$ induced by the
shift-stretch samples is $\approx20\%$ larger than the input samples.  
This mismatch indicates that some cosmologically relevant information in the
original samples is being lost, or that the shift-stretch samples are
imposing cosmological constraints that are not present in the original
samples.

For simplicity, we have demonstrated this method for a case in which
$\vecn$ specifies the $n(z)$ function for a single sample of DES
galaxies.  The method is fully applicable to any set of nuisance
parameters for which we are given a set of samples from $p(\vecn).$
For instance, in the DES Y6 analysis, we use an $\vecn$ that is a
concatenation of the parameters of $n(z)$ for 4 bins of lens source
galaxies.  Since each output sample from the  redshift-estimate
process describes all 4 bins, we know the cross-correlations between
distinct bins' $n(z)$ values, and the compression process described
herein will properly preserve these correlations in the subsequent
cosmological analysis.  In the DES Y6 analysis, we have also extended
the nuisance parameter vector $\vecn$ to include the multiplicative
errors on the shear measurement method, and create a compressed
$\vecu$ that captures correlations between the multiplicative errors
and the $n(z)$ estimates.

\section{Summary}
  We present a linear dimensionality reduction technique that has the
  aim of making it feasible to produce continuous density estimators
  for nuisance parameters that have no evaluable prior, and are characterized only
  from a set of samples in 
  a high-dimensional space $\vecn.$  This is essentially a principle
  components analysis that is adjusted to separate contributions to
  the detectable consequences of $\vecn$ rather than contributions to
  the Euclidean norm of $\vecn.$   This method enables a rigorous
  marginalization over the distribution of $\vecn$ even for
  non-Gaussian distributions, as long as the derivatives of the
  log-likelihood of the data with respect to $\vecn$ are nearly
  constant over the posterior domain.
  The compression method is also robust to any nonlinearities
    in the model and non-Gaussian likelihoods that lie \emph{within} the
    subspace of $\vecn$ defined 
    by the compression, as long as $\vecn$ components normal to this
    subspace do not have detectable effects on the data model.
 The mode-projection technique
  is essentially a method of projecting away irrelevant
  fluctuations in $\vecn.$
The algebra to derive the compression scheme that minimizes the mean shift
  in computed likelihood $\langle  \chi^2\rangle$ makes the assumption that the
  Hessian $\partial^2 \log\likeli / \partial\vecn^2$ is constant in the
  parameter ranges of interest, but in fact the successful application of this
  compression  to inference requires only the weaker assumption that
  fluctuations in $\vecn$ that are normal to the M-dimensional compression plane
  have   negligible impact on $\log\likeli.$

This technique was developed for the case when
  $\vecn$ represents the redshift distributions of galaxies in DES.  Some
  attempts at sampling the posterior $p(\vecq| \vecc)$ while fully respecting
  the prior $p(\vecn)$ as embodied by the samples $\{\vecn_\alpha\}$ have
  proven infeasible, such as concatenating MC's run at every $\vecn_\alpha,$ or the
  nearest-neighbor hypercube embedding of \citet{hyperrank}.  The method of
  \citet{bridle02} that propagates $p(\vecn)$ into an inflated $\covm_c$ relies on
  assumptions of Gaussianity for the nuisances that are not necessarily valid
  here, and precludes some forms of internal consistency checks.
  The mode-projection technique we describe is
  able to do a better job, with
  principled adherence to the prior embodied by
  the input samples.
  Extensive
  use of this method on the 10 different galaxy samples defined for the Year 6
  analysis of DES data is reported by \citet{y6pz, y6maglim} and \citet{y6wz}.

While this new approach to marginalizing over $n(z)$  has minimal consequence for the
  posterior cosmological parameter estimates in the DES Y6 analysis,
  the mode-projection method is better motivated and will produce more
  accurate posteriors in future experiments where the $n(z)$
  prior's uncertainty is a dominant contributor to the cosmological posterior,
  and can be generally applicable to other analyses where critical nuisance parameters
  do not have evaluable priors.

The compression technique is applicable to any inference with a
  high-dimensional prior known only through a set of samples.  In the
  cosmological realm, one case of high-dimensional nuisance
  parameters might  be a binned representation of the absolute response vs
  wavelength $\lambda$ of some imaging bands used in
  constructing Type~Ia supernova Hubble diagrams, as constrained by observations
  of a finite sample of spectrophotometric samples.  Another case might be
  detailed power spectrum $P_\delta(k,z)$ of density fluctuations created by
  gravity plus baryonic forces, as constrained by ensembles or jackknife samples
  of numerical simulations.  This likelihood-aware PCA should be of use to
  inferences in a broad range of fields.
  
  \begin{acknowledgments}
G.M.B. acknowledges support from NSF grant AST-2205808 and DOE award DE-SC0007901.
W. d’A. acknowledges support from the  MICINN projects PID2019-111317GB-C32, PID2022-141079NB-C32 as well as predoctoral program AGAUR-FI ajuts (2024 FI-1 00692) Joan Or\'o.
The project that gave rise to these results received the support of a fellowship
to A. Alarcon from "la Caixa" Foundation (ID 100010434). The fellowship code is
LCF/BQ/PI23/11970028.  We also thank the anonymous referees for pointing out
portions of the paper in need of improvement.

Funding for the DES Projects has been provided by the U.S. Department of Energy, the U.S. National Science Foundation, the Ministry of Science and Education of Spain, 
the Science and Technology Facilities Council of the United Kingdom, the Higher Education Funding Council for England, the National Center for Supercomputing 
Applications at the University of Illinois at Urbana-Champaign, the Kavli Institute of Cosmological Physics at the University of Chicago, 
the Center for Cosmology and Astro-Particle Physics at the Ohio State University,
the Mitchell Institute for Fundamental Physics and Astronomy at Texas A\&M University, Financiadora de Estudos e Projetos, 
Funda{\c c}{\~a}o Carlos Chagas Filho de Amparo {\`a} Pesquisa do Estado do Rio de Janeiro, Conselho Nacional de Desenvolvimento Cient{\'i}fico e Tecnol{\'o}gico and 
the Minist{\'e}rio da Ci{\^e}ncia, Tecnologia e Inova{\c c}{\~a}o, the Deutsche Forschungsgemeinschaft and the Collaborating Institutions in the Dark Energy Survey. 

The Collaborating Institutions are Argonne National Laboratory, the University of California at Santa Cruz, the University of Cambridge, Centro de Investigaciones Energ{\'e}ticas, 
Medioambientales y Tecnol{\'o}gicas-Madrid, the University of Chicago, University College London, the DES-Brazil Consortium, the University of Edinburgh, 
the Eidgen{\"o}ssische Technische Hochschule (ETH) Z{\"u}rich, 
Fermi National Accelerator Laboratory, the University of Illinois at Urbana-Champaign, the Institut de Ci{\`e}ncies de l'Espai (IEEC/CSIC), 
the Institut de F{\'i}sica d'Altes Energies, Lawrence Berkeley National Laboratory, the Ludwig-Maximilians Universit{\"a}t M{\"u}nchen and the associated Excellence Cluster Universe, 
the University of Michigan, NSF NOIRLab, the University of Nottingham, The Ohio State University, the University of Pennsylvania, the University of Portsmouth, 
SLAC National Accelerator Laboratory, Stanford University, the University of Sussex, Texas A\&M University, and the OzDES Membership Consortium.

Based in part on observations at NSF Cerro Tololo Inter-American Observatory at NSF NOIRLab (NOIRLab Prop. ID 2012B-0001; PI: J. Frieman), which is managed by the Association of Universities for Research in Astronomy (AURA) under a cooperative agreement with the National Science Foundation.

The DES data management system is supported by the National Science Foundation under Grant Numbers AST-1138766 and AST-1536171.
The DES participants from Spanish institutions are partially supported by MICINN under grants PID2021-123012, PID2021-128989 PID2022-141079, SEV-2016-0588, CEX2020-001058-M and CEX2020-001007-S, some of which include ERDF funds from the European Union. IFAE is partially funded by the CERCA program of the Generalitat de Catalunya.

We  acknowledge support from the Brazilian Instituto Nacional de Ci\^encia
e Tecnologia (INCT) do e-Universo (CNPq grant 465376/2014-2).

This document was prepared by the DES Collaboration using the resources of the Fermi National Accelerator Laboratory (Fermilab), a U.S. Department of Energy, Office of Science, Office of High Energy Physics HEP User Facility. Fermilab is managed by Fermi Forward Discovery Group, LLC, acting under Contract No. 89243024CSC000002.
\end{acknowledgments}

\textbf{Contributions:}
GB devised the compression method, wrote the relevant code, and wrote the article text.  MT derived the relevant derivative matrices, and Wd'A applied the method to the DES Maglim bin and produced figures.  Authors Wd'A, MT, AA, AA, GG, and BY developed the redshift samples used in the demonstration, integrated the code into DES pipelines, and advised on the implementation and text.
The remaining authors have made contributions to this paper that include, but
are not limited to, the construction of DECam and other aspects of collecting the data; data
processing and calibration; developing broadly used methods, codes, and simulations; running
the pipelines and validation tests; and promoting the science analysis.

\allauthors
\end{document}